# Morphing graphene-based systems for applications: perspectives from simulations

**Tommaso Cavallucci [1,2], Khatuna Kakhiani[2,1], Riccardo Farchioni[2,3], Valentina Tozzini [2,1] ***

[1] NEST-Scuola Normale Superiore and [2] Istituto Nanoscienze del Cnr, Piazza San Silvestro 12, 56127 Pisa, [3]Dipartimento di Fisica, Università di Pisa, Largo Bruno Pontecorvo, 56127 Pisa.

**Abstract** Graphene, the one-atom-thick $sp^2$ hybridized carbon crystal, displays unique electronic, structural and mechanical properties, which promise a large number of interesting applications in diverse high tech fields. Many of these applications require its functionalization, e.g. with substitution of carbon atoms or adhesion of chemical species, creation of defects, modification of structure of morphology, to open electronic band gap to use it in electronics, or to create 3D frameworks for volumetric applications. Understanding the morphology-properties relationship is the first step to efficiently functionalize graphene. Therefore a great theoretical effort has been recently devoted to model graphene in different conditions and with different approaches involving different level of accuracy and resolution. Here we review the modeling approaches to graphene systems, with a special focus on atomistic level methods, but extending our analysis onto coarser scales. We illustrate the methods by means of applications with possible potential impact.

## 1. Introduction

Since its first isolation[1] and especially after the assignment of the Nobel Prize for Physics in 2010 for its discovery, graphene has triggered great expectations. Most of its properties depend on the coincidence of an number of physical-chemical circumstances: three of the four valence electrons of carbon organize in three $sp^2$ hybridized orbitals producing an exactly planar geometry, while the forth contribute to a vast electronic delocalization over the plane, giving great stability to the electronic and geometric structure of the single sheet and very weak inter-layer interaction mediated only by van der Waals (vdW) forces and favoring bidimensionality. In addition, the specific geometry of in-plane bonds generates the highly symmetric honeycomb lattice, responsible for the peculiar band structure with conic geometry at the K points of the Brillouin Zone (BZ). This is responsible for the exotic electronic properties of graphene[2], such as the ultra-relativistic-like high mobility conduction and the possible existence of Majorana fermions[3].

Its – equally exceptional – mechanical properties, have received less attention. The peculiar chemistry and symmetry also brings an extremely large resistance to tensile strain especially considering its bidimensionality (Young modulus of the order of TPa[4], five-fold that of steel), coupled to very low out of plane (bending) rigidity κ (1-2 eV[5],[6],[7]) which is at least one order of magnitude smaller than that expected from materials with comparable in-plane rigidity[8]. This implies an extreme flexibility associated to strength suggesting a vast range of applications in a variety of high-tech fields[9], included those involving exposure to extreme environmental conditions, such as aero-space technologies[10]. Flexibility and bidimensionality also imply the existence of low frequency quadratically dispersive acoustic phonon branches associate to out-of-plane displacement ("flexural" phonons), which can be described as travelling ripples[11] and have a main role in thermal behavior and related phenomena[12].

Many of the applications, however, require graphene morphology manipulation of some kind (see Fig 1). Considering for instance nano-electronics, graphene has null density of carriers at the Fermi level, therefore needing doping to be used as conductor, or band gap opening to be used as semiconductor. These properties can be obtained by escaping from the perfect infinite 2D crystal case, e.g. in multilayers or breaking the honeycomb lattice symmetry. Graphene symmetry disruption is not difficult *per se*, but it is difficult to achieve in a controlled fashion.

Graphene is lightweight and with a huge surface to mass ratio, therefore is considered potentially interesting for gas storage applications. However, it is rather inert, little reactive and with weak vdW interactions. Therefore, this class of applications requires enhancing either physical or chemical interactions. Considering for instance hydrogen storage[13] for energy applications, accurate experimental evaluations of physisorption[14] show that a limiting value for gravimetric (i.e. the relative hydrogen mass stored with respect to the mass system) density in graphene-based systems is ~1% at room temperature (5%-6% at cryogenic temperatures). On the other hand, considering chemisorption one can in principle reach 6-8% gravimetric density at room temperature, but high chemi(de)sorption barriers makes the kinetic of the process very slow at room temperature[13]. Therefore, efforts are directed both to enhance physical-like interactions and to lower the kinetic barriers for chemisorption. Again, manipulation of morphology is required. As for electronics, achievement of interesting properties must proceed through rupture of the perfect graphene symmetry.

While for electronics 2-dimensionality is useful, for other applications the large surface to mass ratio must be declined in a 3D context. This is the case of energy storage applications, including also batteries and supercapacitors, besides hydrogen (or other gases) storage. Therefore great efforts are directed to building 3D graphene based frameworks[15]. This involves including spacers between sheets[16] with tailored size and mechanical properties, and control their amount and distribution[17]. Building 3D networks, controlled chemical functionalization is crucial in a number of other applications[18], including bio-medical ones[19].

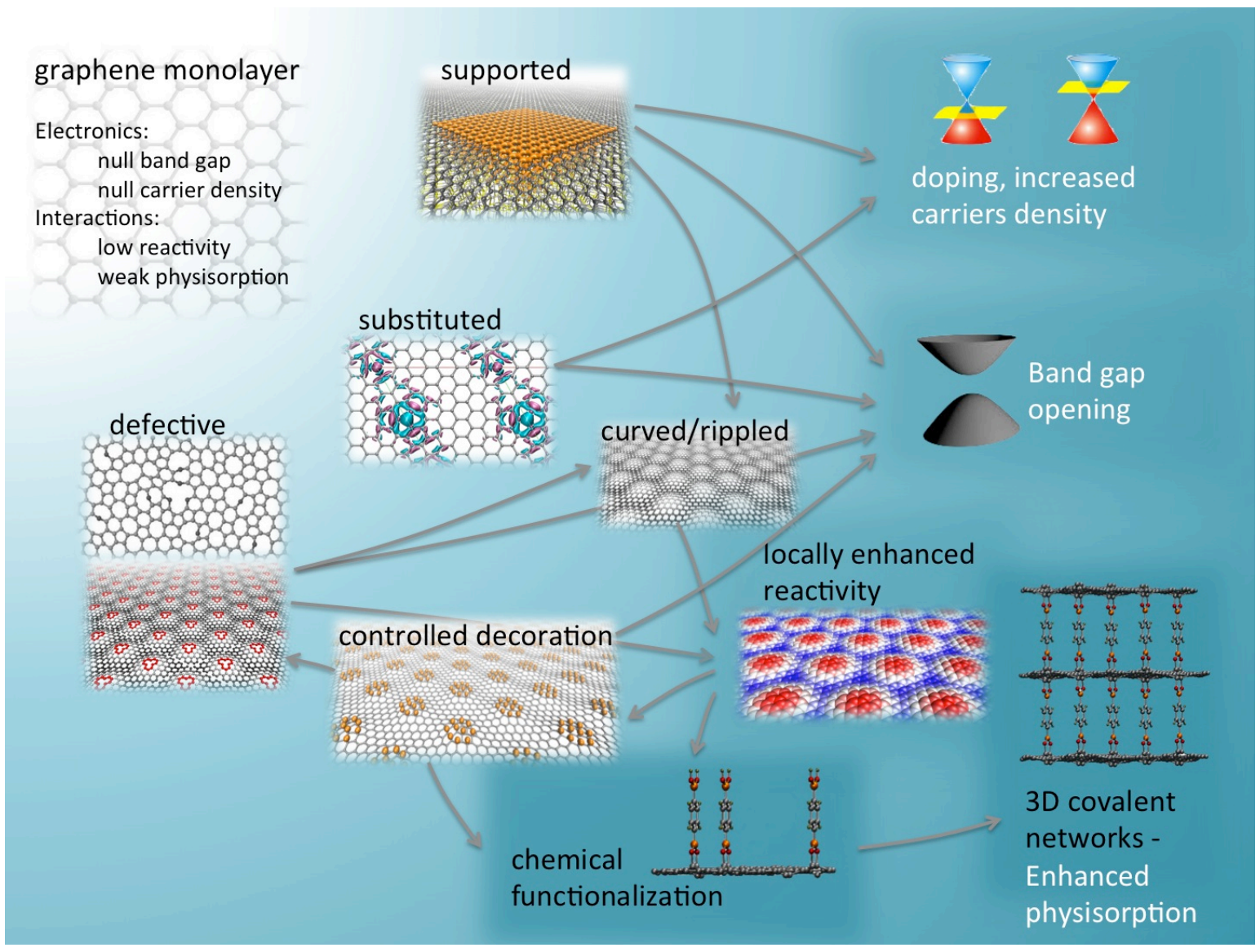

**Figure 1**. A schematic illustration of the relationship between graphene modification and properties in graphene. The bare single layer graphene is located in the upper-left corner. Background increasing level of shade is roughly proportional to the level of manipulation, from simpler ones (inclusion of defects or substitutions, rippling) to the creation of covalent networks. The effects on electronic and chemical properties and physical interactions are reported in white. The arrows indicate the inter-relationships: e.g. the substrate create rippling and in some cases doping and gap opening; chemical substitutions with N and B induce chemical doping and bad gap opening; defects opens the gap and induce curvature; curvature in turn induce local change in reactivity, and consequently controlled decoration and band gap opening. Finally, locally enhanced reactivity could be used to control decoration, chemical functionalization and finally to build 3D covalent network with enhanced physisorption properties. [The images of the chemical functionalization and 3D networks were adapted from ref [16]].

It appears that manipulation of graphene morphology, structure, electronic properties, chemistry and functionalization are interconnected tasks assuming a key role in all emerging graphene related technologies (see Fig 1). In this this context, theoretical investigations and computer modeling are tools of outmost importance to study and possibly design properties of these new materials. The so-called multi-scale approach, traditionally designed and used for biomolecular systems[20], where the hierarchical organization in different length scales is more apparent, is currently being extended to materials[21]. In the cases under discussion, multi-scaling mainly involves the two basic levels of representations, namely the “ab initio” quantum chemistry (QM) level, with explicit electrons capable of describing the chemistry of the system, and the classical “molecular mechanics” (MM) representing the inter-atomic interactions with empirical force fields (FF).

Super-atomic descriptions are also "natural" in specific cases: For instance, fullerenes were sometimes treated with a Coarse Grained sphere-of-beads model[22] similar to that used for biomolecules[23]. For the graphene sheets, conversely, a natural low-resolution representation is a 2D membrane endowed with appropriate mechanical properties[24], analogously to those used for biological membranes.

The scope of this report is to analyze the contribution of modeling to the morphing of graphene-based system for applications. Therefore, for each of the different types of morphing described, the focus is on theory and models able to better describe and predict the properties under examination. The well-known properties of perfect single sheet graphene are also summarized in the next section to have a base for comparison for changes due to manipulation. The subsequent section describes several ways aimed at controlling the electronic properties in graphene-derived systems. The manipulation of both physical and chemical interactions of graphene is addressed in section 4. Section 5 focuses on dynamical deformation of graphene. Section 6 reports a summary and conclusions.

## 2. *Single sheet suspended graphene*

The electronic and structural properties of graphene are very well known and described in a number of very good reviews[2]. The minimal unit cell of graphene (represented in blue in Figure 2a) includes two C atoms. The eight valence electrons organize in four filled bands whose structure was calculated with all available electronic structure methods[25]. The band structure evaluated within the Density Functional Theory (DFT) framework (PBE functional[26]) is reported in Fig 2c (left plot). The $\pi$ band crosses its empty counterpart (in red, the filled bands are in blue) at the K point, where the linear dispersion is described in terms of the Fermi velocity $v_F$. Inclusion of many body effects show on average relatively small corrections to the DFT picture, resulting in a renormalization of $v_F$, whose entity is still under debate[27].

Clearly, when treating an isolated single sheet graphene with no defects or structural deformation, the unit cell is usually the preferred choice for the model system. However, there are infinite ways of defining a graphene supercell. Larger, rotated, or differently shaped cells might be necessary in the presence of substrates or other elements breaking the symmetry of the honeycomb lattice. A selection of them is reported in Figure 2a, while their corresponding Brillouin Zones (BZ) in Figure 2b. The size, shape and orientation of the BZs change accordingly. The definition of different BZs implies that the description of the band structure is cell-dependent. As an example, in Figure 2c we report the comparison of the graphene band structure evaluated in DFT-PBE in unit cell and in the $4\sqrt{3}\times4\sqrt{3}$R30 cell[26,28] (colored in green in Figure 2a,b), including 96 C atoms. Once the scale of energies of the two band structures are aligned, the bands of the $4\sqrt{3}\times4\sqrt{3}$R30 cell appear refolded. In order to understand the refolding, we follow for instance the band along the M-K line of the unit cell, which is fragmented in a sequence of Γ-M-Γ-M-Γ lines of the 4√3×4√3R30 cell, as it can be seen in Figure 2b. The fragmentation is reported in the central plot, and the refolded empty and filled π bands are highlighted in the right band structure. Interestingly, the crossing of bands at K point is remapped onto the Γ point in the 4√3×4√3R30 cell, as an effect of refolding/rotation. It is also interesting to remark that, while the band representation changes, the value of physical observables should not. As a matter of fact the density of electronic states (DoS) evaluated in the two cases coincide (within numerical error) once the energy scales are aligned. This is true in particular for the two peaks at ±1.5eV appearing where the π bands become flat (π edges).

Finally, we remark that the use of rectangular cells allows remapping the main symmetry directions along the two Cartesian directions $x$ and $y$.

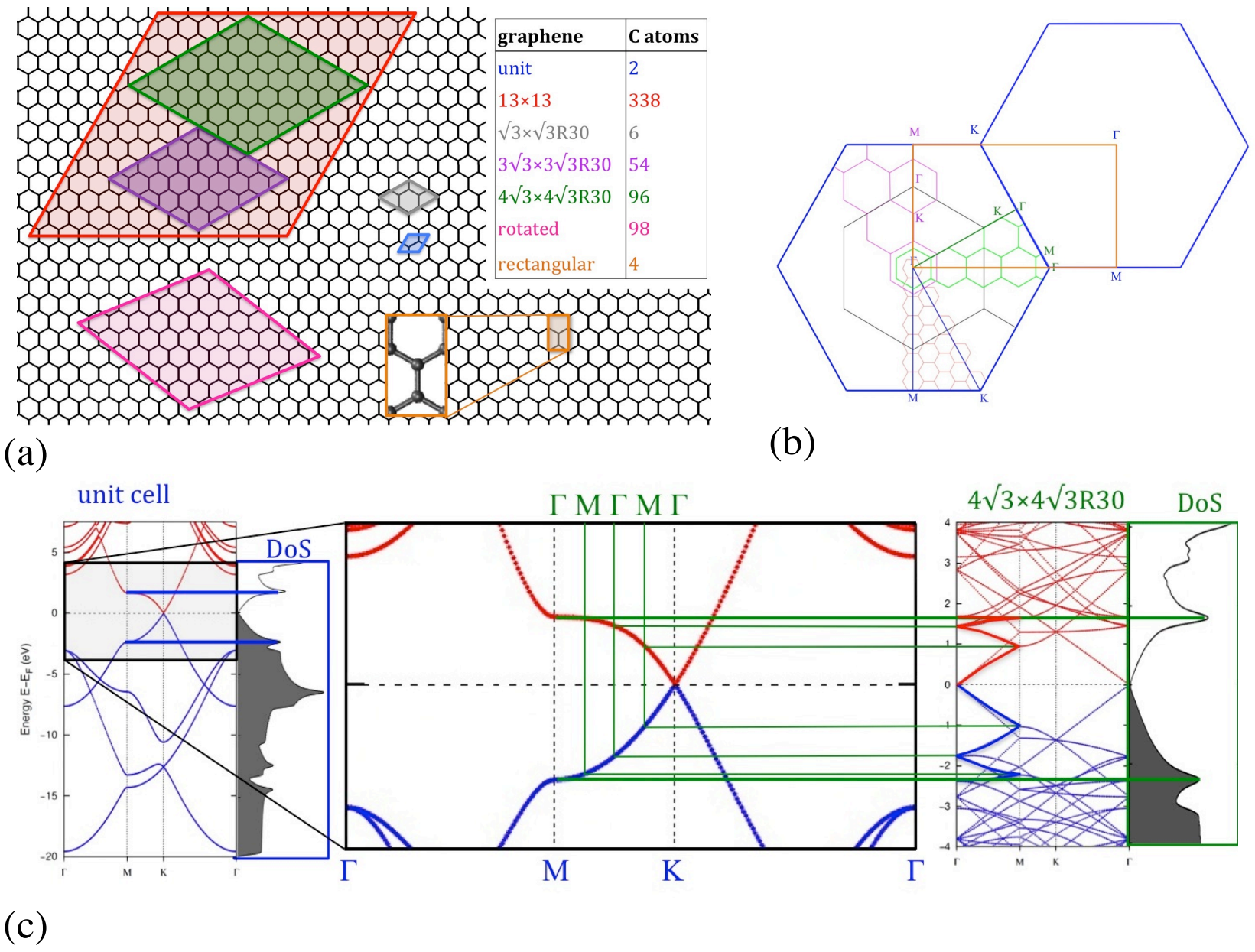

**Figure 2**. Equivalent representation of the graphene structure. (a) Representation of graphene of graphene cells of different sizes, orientation and shape. The number of atoms included is reported. (b) Representation of the BZ of the cells reported in (a) (except the one rotated with 98 atoms) with the main symmetry points. Color coding is the same as in (a) and the symmetry points are colored as the cell BZ they refer to. (c) Electronic structure of single layer flat graphene evaluated within DFT-PBE approach (see text). On the left: Band structure and DoS for the unit cell (in blue in (a)); Center: same expanded in a -4eV – +4eV. When the 4√3×4√3R30 cell is used, the M-K line of unit cell is fragmented in the Γ-M-Γ-M-Γ line (see (b)), as represented by green lines. This produces a refolding of the band, which is represented in the band structure of 4√3×4√3R30 reported on the right (with its DoS). Filled bands are in blue, empty in red. In the DoS, filled states are in grey, empty in white. The edges of the π bands are highlighted with horizontal thick lines (blue in the unit cell bands, green in the 4√3×4√3R30 bands structure).

## 3. *Modification of graphene electronic properties*

Any breaking of the graphene perfect symmetry induces changes in the electronic or transport properties. In the following we review the effects of substrates, chemical substitution/adhesion and structural modifications on the electronic properties.

### *3.1 B and N substitutions and structural defects*

Hexagonal Boron Nitride, BN, has a 2D honeycomb structure identical with that of graphene, with the two triangular sublattices occupied by B and N, respectively. The sublattices symmetry breaking induce the opening of a large band gap, estimated in ~4.5 eV within DFT scheme (generalized gradient approximation[29]), making it the insulating material more similar to graphene. Graphene partially substituted with BN patches or strips of different configuration shows intermediate behavior and can be considered therefore semiconductor systems with band gap tunable as a function of the amount and size of BN areas with respect to graphene. Independent DFT estimates[30,29,26] indicate that the band gap opening is roughly linear at low B/N density, and can be estimated in ~0.03-0.06eV for 1% of substituted C atoms. In presence of excess of N or B, doping (of n-type or p-type respectively) is also present, estimated in ±0.3-0.4eV of Fermi level shift with respect to Dirac point for each % point of N(B) excess[31]. In this case, for given shapes of the substitution patches, mid-gap states and specific magnetic properties can also appear.

Purely structural defects in graphene such as isolated or aligned dislocations delimiting grain boundaries can also produce severe changes in the electronic structure[32]. Depending on the specific periodicity and topology, and on the relative orientation of domains, the system can display conductive behavior or band gaps with reflective transport behavior at the grain boundaries[33]. Vacancy type and Stone-Wales type defects can open band gaps on the scale of eV leading to semiconducting behavior[34] which opens to the possibility of using defects to engineer the electronic properties of graphene. Clearly, due to rupture of the regular $\pi$ electronic system, structural defects of any kind also produce structural deformation and changes in local reactivity, which are discussed in Section 4.

### *3.2 Effects of the substrate*

In supported graphene, the interaction with the substrate can induce electronic properties modifications directly, or indirectly (e.g. by substrate-induced defects) or influence electronic modification due to other effects. Considering graphene grown on SiC by Si evaporation, for instance, it was early inferred by DFT studies that – for grown on the C-rich face, the substrate can enhance the effect of Al, P, N and B substitutional doping[35]. Of all dopants, N is shown to prefer the substituting the graphene upper layer, while other dopants prefer interstitial or buffer layer locations. The enhancement effect seems due to an increased stability of dopants operated by interaction with the substrate. In a model without the buffer layer[36] (whose presence is still debated for grown on the C-exposed surface) a direct effect of the substrate producing n-type doping of ~0.3-0.4 eV, was reported, very sensitive to the intercalation of passivant H atoms. This work also reports a marked sensitivity to the use of different density functionals and treatments of vdW interactions, confirming a main role of the interaction with the substrate. In the Si exposed surface the presence of a covalently bond buffer layer was early established, and calculations with a minimalist cell shows noticeable levels of doping[37].

Early calculations however, were performed in very small simulation cells, inducing strain on the graphene layers to achieve commensuration with the substrate. Subsequent calculations in larger and more relaxed supercells revealed a reduction of all the effects: on the C-rich surface, the first graphene layer becomes almost detached and recover graphene-type metallicity, with little doping[38]. In the case of the Si-rich surface the stress can be completely relaxed only using a very large supercell[39], where doping seems to be negligible. The question is still controversial however, because experimental observations indicate a variable level of n-doping in monolayer graphene, which tends to decrease in quasi free standing graphene (QFSG) obtained intercalating H underneath the buffer layer, depending on the amount of H, and negligible interactions of QFSG with the substrate are confirmed by DFT calculation in the fully relaxed model[40], reporting a ~5-8meV band gap and basically no doping[41]. A more advance theoretical analysis indicate that perturbation from the non-interacting situation and others effects (such as the appearance of specific non linearities in the bands) might be due to many body electron interaction and phonon-coupling effects not included in the standard DFT treatment[42]. Conversely, metallic substrates induce a genuine doping whose origin is ascribed by the difference in work function and to the more or less "physical" or "chemical" interactions between graphene-metal. A systematic DFT based study shows a n-type doping for Al, Ag, Cu, and p-type for Au and Pd[43].

Effects of substrates and of defects can combine. For instance, vacancy type defects created on QFSG are shown to strongly interact with the H coverage and induce doping and/or localized states[44] and metals intercalation (e.g. Li) produces a strong n-type doping. In addition, the procedures to induce substitutional doping, e.g. with N, often induce also vacancy type nitrogenated defects[45] (e.g. with pyridinic or pyrrolic type reconstruction), which combines and often enhances the doping and band gap opening effects, allowing in principle a fine tuning of the electronic properties of the graphene sheet[46].

## *3.3 Effects of adatoms*

Chemisorption of adatoms (typically H and F) is another common way to change graphene properties. Because the fully hydrogenated sheet – graphane – is an insulator[47], partially hydrogenated graphene display semiconducting properties, with band gap depending on the coverage and decoration. Decoration in strips produces the typical ribbon-like behavior[48], with gap dependent on the edge type (zig-zag or harmchair) and decreasing with the width of the hydrogenated strips, and therefore increasing with the H coverage. Decoration in patches produces very variable situations[49,50] depending on the shape and connectivity of the patches, sometimes including midgap non dispersive states[51]. However, when experimental and DFT data are put together as a function of the H coverage, irrespective of the decoration type (Figure 3), they seem to accumulate onto a line which was empirically fitted with an almost square root behavior (precisely, a 0.6 exponential[51]). Best adherence to the empirical curve is observed for the uniform coverage[52] or in general in experimental cases, when the coverage is likely to be more random. Cases of regular or symmetric coverage seem to bring larger dispersion from the curve. Considering the dispersion of the curve, a measurement of the band gap e.g. with STS techniques could give an evaluation of H coverage with an error of 10-15%.

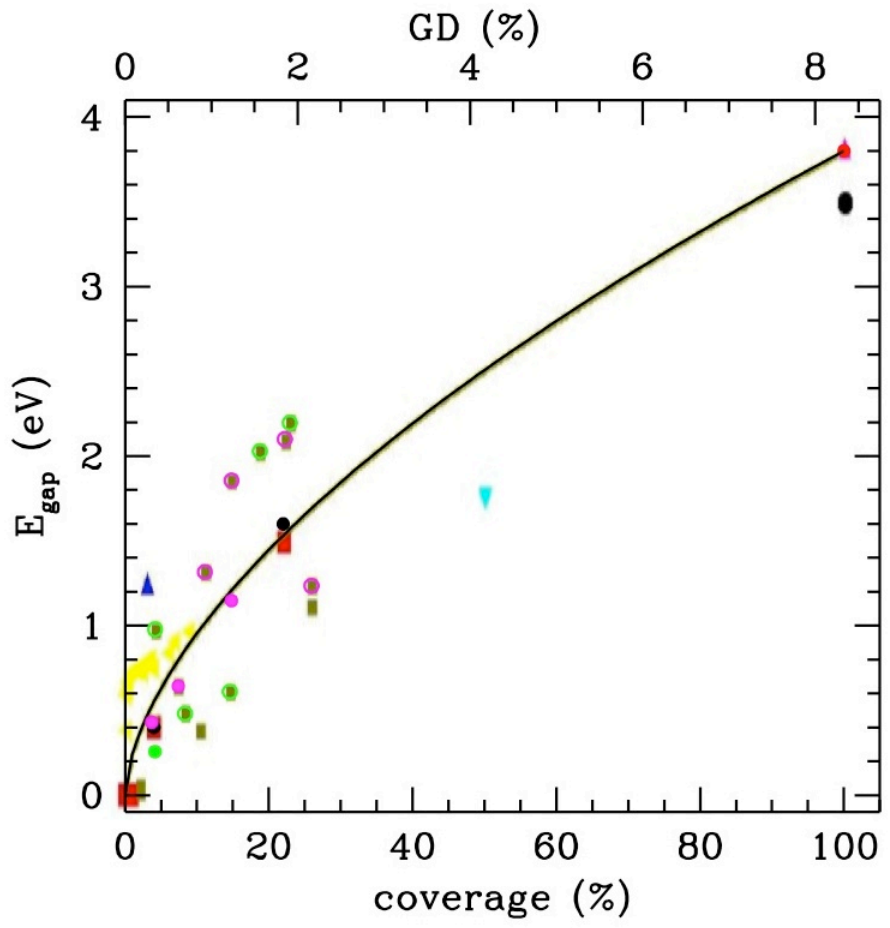

**Figure 3**. Band with of partially hydrogenated graphene as a function of the H coverage. Data coding: line and magenta/green/brown circles/dots are taken from ref [48] (DFT); black dots are from reference 45 (DFT); red squares are from ref [47] (expt and DFT); yellow data are from ref [50] (DFT and expt); blue and cyan triangles from ref [49] (DFT).

Due to its symmetry, single layer graphene is not piezoelectric, although there are indications of its reactivity to electric fields by means of flexo-electric behavior[53,54]. Again, the sheets functionalization with specific elements (H, F, Li, K) was shown an effective way to induce piezo-electricity: fluorination and decoration with Lithium, in particular, are shown to bring piezoelectric coefficients of the order of magnitude of those of 3D matherials[55]. Piezo-electricity can also be induced creating holes with the right symmetry on the layer[56], again breaking the graphene inversion symmetry. Chemical manipulation and substitution also influences the flexo-electric properties: curvature changes induced by an external electrostatic field are enhanced in the presence of N substitutions[26] (Figure 4).

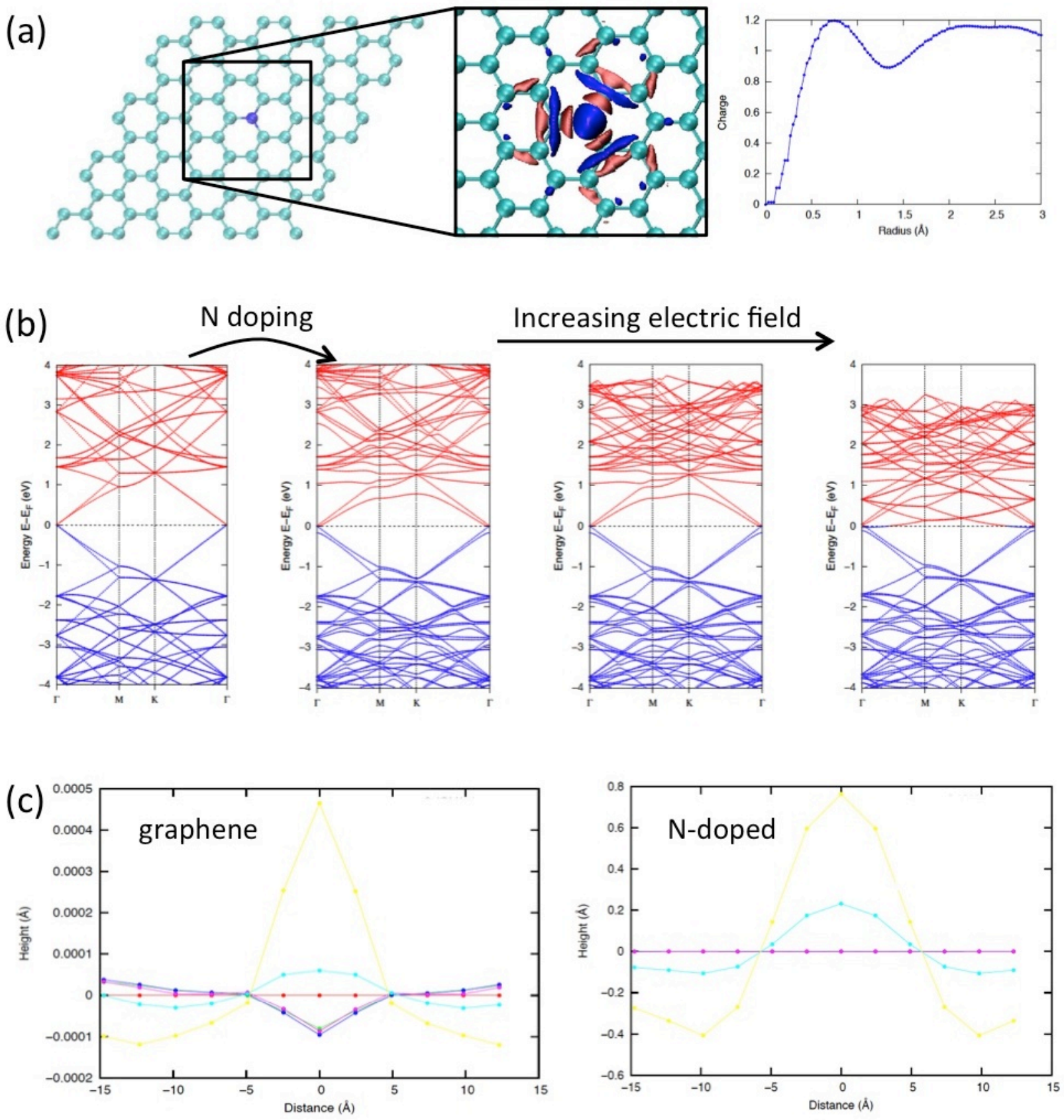

**Figure 4**. Effect of N-doping on flexo-electricity of graphene (adapted from ref [26]). (a) 4√3×4√3R30 graphene supercell with N-substitution in the center. A detail of the charge redistribution is reported (blue=charge accumulation, pink=charge depletion) and the radial distribution of electronic charge is also reported in the plot. (b) Change of the band structure as an effect of doping and of increasing electric field applied orthogonally to the sheet (max field value = 10GV/m). Besides the small band gap opening and doping level due to substitution, electric field bends the bands and enhances doping. (c) Height profile along the main diagonal of the supercell, for bare graphene (left) and N-doped graphene (right) at increasing levels of the electric field. In bare graphene the effect is very small and fluctuating also in directions. In N-doped graphene the effect is 3-4 orders of magnitude enhanced, due to electronic charge localization and symmetry breaking. A systematic study of the effects of doping on flexoelectricity is currently in the course.

Table 1 reports a summary of the cited DFT based works on graphene with different morphological modifications.

| System | Calculation | Main results | Ref |
|---|---|---|---|
| BN substituted single sheet | 6×6 supercell (72 atoms) GGA functional | Band structure, electronic gap and dos variation as a function of the BN relative amount and decoration shapes | [29] (2011) |
| Mono and Bi-layers with B or N substitutions | 4×4 supercell (32-64 atoms) PBE functional | Doping level for localized substitutions of B or N | [30] (2009) |
| Monolayer with BN patches | 6×6 supercell (192 atoms) PBE functional | Band gap and doping for different size/shape of the patches and unbalance between B and N | [31] (2010) |
| Monolayer with dislocations and grain boundaries | Rectangular supercell of ~4×1.5 nm size (estimated ~150 atoms) GGA spin polarized | Electronic and transport properties dependence on the topology of defects and relative orientation of grain boundaries | [32] (2010) |
| Monolayer with "octite" defects isolated or in super-lattices | Rectangular supercell, (144–200 atoms) DFT and GW | Band gap in different configuration is evaluated and shows marked dependence on the distribution and symmetry | [34] (2010) |
| Graphene on SiC (C-rich surface) | 4×4 supercell, with buffer layer and substrate (~150 atoms); LSDA | Band structure, evaluation of preferential subsitutions and level of doping of different dopants (P, Al, B, N). | [35] (2011) |
| Graphene on SiC (C-rich surface) | 2×2 without buffer layer PBE+ empirical vdW corrections | n-type doping is naturally induced by the intearction with the substrate, but dependent on the type of passivation (with or without intercaled H), and on the inclusion of the correct vdW interactions. | [36] (2011) |
| Graphene on SiC (Si-Rich surface) | 2×2 with buffer layer and substrate GGA | n-type doping of 0.5eV is observed, induced by the buffer-layer – monolayer interaction | [37] (2007) |
| Graphene on SiC (C-rich surface) | 5×5 with buffer layer and substrate DFT | First layer is almost metallic, with little doping. | [38] (2011) |
| Graphene on SiC (Si-rich surface) | 13×13 with buffer layer and substrate; LDA | The buffer is covalently bound, the first monolayer is metallic and with negligible doping | [39] (2008) |
| Quasi Free Standing on SiC and Li intercaled graphene | 13×13 with complete intercaled H coverage; PBE+vdW | The QFSG is weakly interacting, with very small band gap and basically no doping | [40] (2011) |
| Quasi free standing graphene on SiC | 13×13 with buffer layer and substrate and complete intercaled H coverage; PBE+vdW | The interaction between substrate and graphene is very weak, leading uniform electronic density on rather flat sheet | [42] (2015) |
| Graphene on different metals (Cu, Ni, Co, Pt, Pd) | Unit and 2×2 cells, unstreched, with metal lattice adapted accordingly; LSDA | Graphene interaction with different metal has different character (physical or chemical), and different type and level of doping | [43] (2008) |
| Quasi free standing graphene on SiC with vacancy type defects | 13×13 with buffer layer and substrate and complete intercaled H coverage; PBE+vdW | The defect strongly interacts with the substrate, producing changes in eleand magnetic ctronic properties | [44] (2014) |
| Single sheet with N substitutional and pyrrolic/pyridinic defects | 3×13 with N substitutions and/or vacancies; LDA | The effects of substitution and vacancy allow a tuning of the band gap and doping | [46] (2011) |
| Single sheets | Up to 400 atoms, rectangular geometry; LDA | Evaluationof flexoelectric coefficient | [53] (2008) |
| Single sheets fluorinated, hydrogenated , fluorinated and decorated with Li and K | Unit, 2×2, 3×3, 4×4 cells PBE | Evaluation of piezoelectric coefficient in functionalized graphene | [54] (2012) |
| Single sheet, with saturated vacancy defects | Rectangular (10×5) | Evaluation of piezoelectric coefficient in graphene with vacancies | [56] (2012) |

**Table 1**. As survey of the modeling works on (morphed) graphene systems. Density Functional Theory based calculations and simulations.

In conclusion to this section, we remark that while standard DFT methods seems appropriate to evaluate the energetics and electronic properties for chemical manipulations, the effect of substrates, especially when the interaction is more physical in nature, calls into play the use of more advanced DFT schemes, involving the use of vdW corrections and evaluation of many body electronic effects.

## *4. Reactivity and interactions manipulation*

From previous discussion the necessity of controlling adsorption or substitution of chemical species onto graphene emerges as a way to endow graphene with electronic and electro-mechanical properties of outmost interest in applications. Controlling interaction of graphene with atomic or molecular species has also a direct use in energy applications: graphene has a huge surface to mass ratio, therefore it would be in principle an ideal candidate not only for gas (mainly hydrogen) storage applications, but also for super-capacitors and batteries[57]. In addition, reactivity control is a key step for building 3D networks with organic inter-layer spacers. In the following a selection of theoretical and simulation works on adsorption of gases (mainly hydrogen) in graphene based system is reported. A summary of the literature is in Table 2.

### *4.1 Physisorption*

Taking hydrogen as a paradigmatic case for molecular interactions with graphene, one can generally separate adsorption onto graphene in two main classes, chemisorption and physisorption[13]. Physisorption occurs barrierless via dispersive vdW interactions in case of neutral molecules. In the case of $H_2$, the binding energy on bare graphene is estimated around 0.01eV[58], bringing a precise linear dependence of the gravimetric density onto the specific surface area[14] and posing a strict upper limit to the gravimetric density for physisorption onto graphene at 1-2% at room temperature. Besides the obvious strategy of working at cryogenic temperature, in several theoretical works physical interactions were shown enhanced within nanocavities produced by rippling[58], within multilayers with specific (nano-metric) interlayer spacing[59,60] or with nano-sized perforations of the sheet[61].

From the methodological point of view, it is interesting to observe that these works address the problem of physisorption combining different methodologies besides DFT (see Table 2), including classical approaches based on empirical Force Fields, different types of dynamical treatment of hydrogen (QM dynamics, Monte Carlo and thermodynamic evaluation of the adsorption). This is basically due to two circumstances: first, gravimetric density evaluation must be treated at the statistical level, which requires large system and sampling methods proper to this aim. Second the dispersive interactions are very elusive and difficult to treat with standard DFT methods. Commonly used LDA and GGA Density Functionals are based on the (semi)local representation of exchange-correlation, insufficient to address instantaneous fluctuations of the electron density. As a consequence, these functionals fail in reproducing the medium-long range behavior of inter-atomic interactions[62,63], which was corrected with a number of semiempirical schemes. In addition, hydrogen, the lightest element, calls a quantum treatment of its dynamics. Because all of these circumstances make the simulation unaffordable on large systems, recursion to combination of high accuracy/high cost with lower accuracy/low cost methods in multi-scale combined approach is compulsory. Overall, however, the lesson one learns from this analysis is that cavities, convexities, porosity or in general structures enhance the physisorption at best when they are at the nano-scale (specifically ~1nm). Therefore nano-scale structuring must be considered a guideline also in designing graphene-based 3D networks.

## *4.2 Manipulation of reactivity*

In contrast to physisorption, chemisorption on graphene is barrier driven process, requiring a change in hybridization from $sp^2$ to $sp^3$ [64, 65]. The barrier for atomic H is estimated around 0.3eV, but the barrier for $H_2$, involving dissociative chemisorption, is ~1.5eV per atom. Once adsorbed H can hop from site to site with a barrier of ~0.75eV[66]. These processes display slow kinetic at room temperature.

Enhancement of reactivity can be obtained by perturbing the $\pi$ delocalization in several ways, which mostly superimpose to those already mentioned to manipulate electronic structure. Structural and substitutional[67,68,69,70] defects generally constitutes hot-spots of reactivity. The action mechanism is usually ascribed to the presence of localized states or doping due to the defect, which are able to induce dissociation of the molecule and subsequent adsorption of the reactive radicals, resulting in a reduction of the dissociative chemisorption barrier. External electric fields were shown to act as catalysts for this process[71,72], being able to deform the already perturbed orbitals. This effect is synergetic with that of substitutional doping[73,74,75] promising fast kinetics, large GD (6.73%) at room temperature and the possibility of using electric field for switch uptake/release of molecules.

The reactivity enhancement effect of purely structural defects is partially due to the local distortion from flatness, which induces a protruding and partial pyramidalization of a C site, favoring the $sp^2$ to $sp^3$ transition. In fact, significant lowering of the barriers for $H_2$ dissociation were previously observed on nanotubes[76], while stabilization of the adsorbate was observed on corrugated graphene without defects[77,78,58]. This indicate that control of local curvature can induce control of graphene decoration, upload and even release, as demonstrated by a simulation showing $H_2$ release by curvature inversion[58]. A side (but related) effect, is that the band gap also depends on mechanical deformation, therefore rippling[28] or stretching[79] could be used directly to manipulate electronic properties, and indirectly by curvature induced chemical functionalization.

Because chemisorption involves much larger energies than physisorption, in a first approximation vdW correction to standard DFT schemes play a less important role. However they become crucial in the approaching phase of the molecule, when different orientations and locations with respect to C atoms determine the effective reaction path and barrier for the chemisorption reaction[80,81,82].

## *4.2 Metal mediated adsorption*

Metal mediated adsorption of hydrogen is often located[83], in between chemisorption and physisorption. This interaction is specifically mediated by d orbitals in transition (e.g. Ti[84]) or heavy metals (e.g. Pd[85],[86] which also works as catalyst and dissociate the molecule producing a "spillover" effect of atomic H onto graphene), while it is often described by an "enhanced" vdW interaction for lighter metals such as Li and Ca[87]. In both cases the interaction energy is increased of at least one order of magnitude, which suggests useful applications in storage. However the problem is turned in how to control the distribution of metal onto graphene. In fact in most cases metals tend to form clusters, reducing the active surface for adsorption with respect to total mass.

Again, combination of different kinds of graphene manipulation has a key role. For instance, it was shown that inducing N-defects onto graphene before exposure to Ti can reduce the cluster size, optimizing the active surface for adsorption[88]. Combination of Li decoration, N-doping and electric field predict reversibility of hydrogenation in a DFT study[89].

| System | Calculation | Main results | Ref |
|---|---|---|---|
| Single layer and multi-layer, rippled | Rectangular cell, 180 atoms; PBE | H chemisorbs on convexities and is unstable within concavities. $H_2$ physisorbs preferentially within concavities | [58] (2011) |
| Multilayer with different layer spacing, with molecular hydrogen | 30×30 supercell; Accurate H2-graphene potential fitted on post-Hartree-Fock theories; Quantum treatment of $H_2$ dynamics | For specific inter-layer spacing, the hydrogen adsorption is enhanced | [59] (2005) |
| Multilayer graphene-oxyde framework pillared with diboronic-derived molecules exposed to $CO_2$ | ~2nm×2nm×2nm<br>DFT-PBE, DFTB<br>Classical MMFF94 | Using realistic framework models with the correct spacing and corrugation enhances selectivity and the adsorption coefficient. | [60] (2015) |
| Multilayer perforated graphene exposed to H2 | Nano-sized supercells. Classical LJ interactions for H2-gr interaction<br>Grand Canonical Monte Carlo dynamics for hydrogen | Graphene perforation effectively enhances the active surface and increase the physisorption capability | [61] (2015) |
| Sheets with H, O, OH, $CH_3$ ad-groups | 30×30 supercell; GGA | Evaluation of binding energies and hopping barriers | [66](2009) |
| Sigle sheet exposed to H2 embedded in electric fields | 2×2 supercell; LDA | Barriers are lowered by EF orthogonal to the sheet | [69] (2010) |
| Single sheet exposed to $H_2$, N-doped, embedded in EF | 4×4 supercell; RPBE | The effect of N-doping and EF cooperate, leading to substantial decrease the chemisorption barrier | [75] (2012) |
| Sheets corrugated with sinusoildal out of plane deformations of different wavelength | Rectangular supercells 28 to 84 atoms; PBE | The fluorination and hydrogenation energy depends on local curvature | [78] (2011) |

**Table 2.** A survey of literature on chemi and physisorption of $H_2$ onto and in graphene

Morphology manipulations have diverse effect on selectivity of adsorption process of molecules on graphene, so that it can be used for the gas sensor applications. Doping graphene with heteroatoms can results in different amount of valence electrons, and usually protrude (i.e. dopants Al, Si, P, S, Mn, Cr)[90,91] from the graphene surface and create reactive binding sites for molecular adsorption. In general, as in the case of physisorption and more than for chemisorption, the vdW interaction have a main role, therefore the use of appropriate functional for their representation is crucial[92].

## 5. Dynamical morphing

Reversibility of graphene manipulation is a key property in tasks such as hydrogen storage. Mechanical deformations are the most reversible. In fact the reversible storage using the reactivity dependence on local curvature was demonstrated in simulations[58]. Therefore, the problem is turned on how to control local curvature of the graphene sheet. One possibility is to exploit the natural curvature of supported graphene: on given substrates, such as SiC, graphene shows multi-stable patterns of curvature[93], allowing the possibility of switching between each other by changing external environmental conditions (e.g. temperature, electric fields).

However, graphene offers the unique possibility of changing the local curvature dynamically, exploiting the naturally occurring out of plane wave-like deformations, i.e. the flexural phonons (or acoustic z polarized branches, i.e. ZA phonons). The capability of those vibrations to chemisorbed hydrogen and transport and pump it though a multilayer system was demonstrated by DFT[58] and classical MD simulations based on empirical FFs[11]. Clearly, for such applications, the key issue is how to create and maintain coherent ZA phonons of specific wavelength and amplitude.

Many efforts were devoted to the calculation the phonon dispersion curves of graphene. The reason is that due to the 2d nature of this material, the vibrational properties play a crucial role on its thermal and mechanical properties. Much attention has been reserved to the ZA mode, particularly important in 2d materials as graphene[94]. Its dispersion law is parabolic, implying that these modes are the lowest energies of the spectrum, and are the easiest to be excited. The very low phonon-phonon scattering rate and the large thermal population have as a consequence that the flexural phonons give a fundamental contribution to thermal conductivity both for graphene monolayer[95] and multilayers[96]. Moreover, they are responsible of the negative thermal expansion coefficient observed in a wide temperature interval, up to an inversion temperature value, which is still controversial.

| **System** | **Calculation** | **Main results** | **Ref** |
|---|---|---|---|
| Suspended graphene sheet | 0.4x20nm supercell AIREBO FF | Suspended graphene deflects in the presence of an electric field orthogonal to the sheet | [51] (2010) |
| Graphene single layer | MD with classical force field | Dynamics of flexural modes and its relationship with thermal and mechanical properties | [95] (2015) |
| Graphene single layer | MD with classical force field | Lattice thermal conductivity of graphene is dominated by contributions from the flexural phonon mode | [96] (2010) |
| Graphene multilayer and graphite | MD with classical force field | Lattice thermal conductivity of graphene is dominated by ZA mode with a lower thermal conductivity due to a break of a selection rule on phonon-phonon scattering | [97] (2011) |
| Graphene nanoribbon | 2d continuum model | Acoustic phonon dispersion curves in the case of fixed and free boundaries | [98] (2011) |
| Graphene nanoribbon | Elastic continuum model | Optical phonon dispersion curves | [99] (2009) |
| Graphene single layer | DFT-LDA, plane waves basis set. rectangular supercell of 128 atoms | Phonon dispersion relations | [100] (2003) |
| Graphene single layer | DFT and DFPT, PBE | Structural, dynamical, and thermodynamic properties, phonon dispersion curves. | [101] (2005) |
| Graphene monolayer, bilayer, trilayer | DFPT | Phonon dispersion curves; optical phonon $E2g$ mode at $\Gamma$ splits into two and three doubly degenerate branches for bilayer and trilayer graphene | [102] (2008) |
| Graphene monolayer | MD with empirical potentials (Tersoff, Tersoff-Lindsay, LCBOP, AIREBO). 20x20 unit cells | Phonon dspersion curves and modification with temperature. Tersoff potential with modified parameterization shows the most physically sound behaviour | [103] (2015) |
| Graphene monolayer and graphite | Standard lattice dynamics with LCBOPII empiric interatomic potential | Phonon dispersion and bending rigidity | [104] (2011) |
| Graphene monolayer | Tersoff and Brenner potential | Optimization of and modification of FF for the accurate reproduction of phonons | [105] (2010) |
| Graphene monolayer | Tersoff potential, parameters in [11] | Calculation of thermal conductivity | [106] (2012) |
| Graphene monolayer | Monte Carlo with empiric interatomic potentials. isothermal–isobaric ensemble, periodic boundary conditions along x,y, reparameterized long-range carbon bond-order potential LBOP and LCBOP | Thermodinamic properties; empirical potentials limited to nearest-neighbour interactions give rather dispersed results. | [107] (2014) |
| Graphene monolayer | Brenner (first generation) | Analytic form for the bending modulus | [108] (2009) |
| Graphene monolayer and single-wall nanotube | Analytic from interatomic potential | tension and bending rigidity directly from the interatomic potential. | [109] (2010) |
| Graphene monolayer | Continuum approach; Molecular mechanics, REBO interatomic empirical potential | Nonlinear elastic properties under uniaxial stretch and tension | [110] (2012) |
| Graphene layer on Si terminated SIC | environment-dependent interatomic empirical potential (EDIP) Graphene layer basic cell of 338 atoms, | Determination of the atomistic structure of the graphene buffer layer on Si-terminated SiC. The solution of minimal energy forms a hexagonal pattern composed of stuck regions separated by unbonded rods that release the misfit with the SiC surface | [111] (2010) |
| Graphene layer | Molecular dynamics simulation, Tersoff empirical interatomic potentials for C–C, Si–Si, and Si–C interactions | Two layers of carbon atoms of the 6*H*-SiC (0001) subsurface after sublimation of Si atoms undergo a transformation from a diamondlike phase to a graphenelike structure at annealing temperature above 1500 K | [112] (2008) |

**Table 3**. Vibrational mechanical and thermal properties of graphene systems, theoretical approaches

Among the most used computational approaches we can mention the elastic continuum model[97],[98], the first principles DFT also with Perdew-Burke-Ernzerhof generalized gradient approximation[99],[100],[101], and the molecular dynamics simulations associated with the use of empirical interatomic potentials for the interactions between carbon atoms, which allows to treat larger systems[102]. The most effective have been demonstrated the Long-range Carbon Bond Order Potential (LCBOPII)[103] and the parameterization by Lindsay et al.[104] of the Tersoff potential[105], which has been also used to calculate the thermal conductivity of graphene[106],[107]. Empirical interatomic potentials have been widely used also for the investigation of the mechanical and elastic properties of graphene sheets, in particular for the elastic bending modulus[108], the thickness and Young's modulus[109],[110], also on Si-terminated SiC[111] and 6H-SiC(0001)[112].

## 6. Conclusions and Perspectives

Morphology manipulation of graphene requires flexible experimental and modeling approaches. In particular on the modeling side, both accuracy in the representation of the phenomena, and the use of model systems very large compared to the atomic scale are required. The former calls for the use of at least DFT based schemes, most often corrected to account for long range electron correlation effects to better account for specific electronic properties or for the dispersive part of vdW interaction; in specific cases the recursion to quantum mechanical treatment of nuclei is also required. On the other hand, large model systems are needed to analyze systems at the nano-micro scale with defects and functionalization, which have a low level of symmetry. The evaluation of thermodynamic properties is also often needed, requiring extensively long simulations. This is often incompatible with the high computational cost of accurate methods, therefore the recursion to empirical treatment of interactions is used. In some cases the two methods, *ab initio* and empirical, are mixed at several levels, intrinsically (such as in the empirical vdW corrections to DFT) or explicitly.

*Ab initio* level currently allows addressing the $10^5$ atoms scale (corresponding approximately to the 100nm size in graphene) in short simulations. Considering the exponential increase of computer power ensured by Moore's law, the μm scale can be reached in less than 10 years, while it is already feasible, in principle, using empirical approaches on computing systems with extensive parallelization.

However, as the scales feasible in simulations increase, new questions emerge on the reliability of the theories underlying the simulation. This is especially true for more empirical approaches, because the parameterization of interactions are usually tested on smaller/shorter size/time scales, but also for more *ab initio* approaches such as DFT, which always include some hidden level of empiricism. Other sources of "systematic" errors in modeling might rise in the algorithms used for sampling of the conformations and phase space of the system and in the large scale average or thermodynamic properties.

In conclusion, as the size of the model system increases, the comparison with measurements becomes more and more important. The sizes of experiment and of simulations tend to meet – the first coming from bottom, the second from top – at the meso-scale. This allows a direct comparison of measured observable with the corresponding evaluated quantity. Clearly this brings an advantage in the interpretation of experiments, but also a feedback on the theory, allowing creating models more adherent to reality.

## Acknowledgments

We gratefully acknowledge financial support by funding from the European Union's Horizon 2020: the Marie Sklodowska-Curie grant agreement No 657070 and the graphene Core 1, Grant Agreement No. 696656; the CINECA award ISCRA-C: Electro-mechanical manipulation of graphene reactivity EIMaGRe, 2015, "ISCRA C" IscraC_HBG, 2013 and PRACE "Tier0" award Pra07_1544 for resources on FERM (IBM Blue Gene/ Q@CINECA, Bologna Italy), and the CINECA staff for technical support. We thank Dr Vittorio Pellegrini and Prof P. Giannozzi for useful discussions

[10] Siochi E. J. 2014 Graphene in the sky and beyond *Nature Nanotechnology* 9:745-747

[11] Camiola V. D., Farchioni R., Pellegrini V., and Tozzini V. 2015 Hydrogen transport within graphene multilayers by means of flexural phonons *2D Materials* 2(1):014009

[12] De Andres P. L., Guinea F., and Katsnelson M. I. 2012 Bending modes, anharmonic effects, and thermal expansion coefficient in single-layer and multilayer graphene *Phys. Rev. B* 86:144103

[13] Tozzini V., and Pellegrini V. 2013 Prospects for hydrogen storage in graphene P*CCP* 15(1):80-89

[14] Klechikov A., Mercier G., Sharifi T., Baburin I. A., Seifert G., and Talyzin A.V. *Chem. Commun.* 2015 Hydrogen storage in high surface area graphene scaffolds 51:15280-15283

[15] Wang H., Yuan X., Zeng G., Wu Y., Liu Y., Jiang Q., and Gu S. 2015 Three dimensional graphene based materials: Synthesis and applications from energy storage and conversion to electrochemical sensor and environmental remediation *Advances in Colloid and Interface Science* 221:41–59

[16] Mercier G., Klechikov A., Hedenström M., Johnels D., Baburin I.A., Seifert G., Mysyk R., and Talyzin A. V. 2015 Porous Graphene Oxide/Diboronic Acid Materials: Structure and Hydrogen Sorption *J. Phys. Chem. C* 2015 119(49):27179–27191

[17] Xia Z., Leonardi F., Gobbi M., Liu Y., Bellani V., Liscio A., Kovtun, Li R., Feng X., Orgiu E., Samori P., Treossi E., and Palermo V. 2016 Electrochemical Functionalization of Graphene at the Nanoscale with Self-Assembling Diazonium Salts *ACS Nano*, Just Accepted Manuscript DOI: 10.1021/acsnano.6b03278

[18] Jiang H., Lee P. S., and Li C. 2013 3D carbon based nanostructures for advanced supercapacitors *Energy Environ. Sci.* 6:41

[19] Georgakilas V., Spinato C, Ménard-Moyon C., and Bianco A. 2014 Chemical Functionalization of Graphene for Biomedical Applications *in "Functionalization of Graphene"* Wiley on-line library DOI: 10.1002/9783527672790.ch4

[20] Tozzini V., 2010 Multiscale modeling of proteins *Accounts of chemical research* 43 (2):220-230

[21] Camiola V.D., Farchioni R., Cavallucci T., Rossi A., Pellegrini V., and Tozzini V. 2015 Hydrogen storage in rippled graphene: perspectives from multi-scale simulations *Frontiers in Materials* 2:00003

[22] Akshay S., Srikanth B., Kumar A., and Dasmahapatra A. K. 2015 Coarse-grain Molecular Dynamics Study of Fullerene Transport across a Cell Membrane *J Chem. Phys.* 143(2)

[23] Trovato F., and Tozzini V. 2015 Diffusion within the cytoplasm: a mesoscale model of interacting macromolecules *Biophysical journal* 107(11):2579-2591)

[24] Jomehzadeh E., and Pugno N.M. 2015 Bending stiffening of graphene and other 2D materials via controlled rippling *Composites Part B* 83:194-202

[25] Konschuh S., Gmitra M., and Fabian J. 2010 Tight-binding theory of the spin-orbit coupling in graphene *Phys. Rev. B* 82:245412

[26] Cavallucci T., 2014 Density Functional Theory simulations of the elctromechanical properties of naturally corrugated epitaxial graphene Master Thesis in Physics University of Pisa.

[27] Trevisanutto P. E., Giorgetti C., Reining L., Ladisa M., and Olevano V. 2008 Ab Initio GW Many-Body Effects in Graphene *Phys Rev Lett* 101:226405

[28] Rossi A., 2014 Corrugated graphene hydrogenation with Density Functional Theory based simulations Master Thesis in Physics University of Pisa.

[29] Shinde P. P., and V. 2011 Kumar, Direct band gap opening in graphene by BN doping: Ab initio calculations *Phys Rev B* 84:125401

[30] Panchakarla L. S., K. S. Subrahmanyam, S. K. Saha, Achutharao Govindaraj, H. R. Krishnamurthy, U. V. Waghmare, and C. N. R. Rao 2009 Synthesis, Structure, and Properties of Boron- and Nitrogen- Doped Graphene *Adv. Mater.* 21:4726–4730

[31] Xu B , Y. H. Lu , Y. P. Feng , and J. Y. Lin J 2010 Density functional theory study of BN-doped graphene superlattice: Role of geometrical shape and size J App Phys 108:073711

[32] Yazyev O. V., and Louie S. G. 2010 Topological defects in graphene: Dislocations and grain boundaries *Phys Rev B* 81:195420

[33] Yazyev O. V., and Louie S. G. 2010 Electronic transport in polycrystalline graphene *Nature Materials* 9(10):806-809

[34] Appelhans D. J., Carr L. D., and Lusk M. T. 2010 Embedded Ribbons of Graphene Allotropes: An Extended Defect Perspective *New Journal of Physics* 12(12):125006

[35] Huang B., Xiang H. J., and Wei S. H. 2011 Controlling doping in graphene through a SiC substrate: A first-principles study. *Phys Rev B* 83(16): 161405

[36] Jayasekera T., Xu S., Kim K. W., and Nardelli M. B. 2011 Electronic properties of the graphene/6H-SiC $(000\bar{1})$ interface: A first-principles study *Phys Rev B* 84(3):035442

[37] Varchon F., Feng R., Hass J., Li X., Nguyen B. N., Naud C., Mallet P., Veuillen J.-Y., Berger C., Conrad E. H., and Magaud L. 2007 Electronic structure of epitaxial graphene layers on SiC: effect of the substrate. *Physical review letters* 99(12):126805

[38] Deretzis I., and La Magna A. 2011 Single-layer metallicity and interface magnetism of epitaxial graphene on SiC $(000\bar{1})$ *Applied Physics Letters* 98(2):3113

[39] Kim S., Ihm J., Choi H. J., and Son Y. W. 2008 Origin of anomalous electronic structures of epitaxial graphene on silicon carbide *Physical review letters* 100(17): 176802

[40] Sforzini J., Nemec L., Denig T., Stadtmüller B., Lee T. L., Kumpf C., Soubatch S., Starke U., Rinke P., Blum V., Bocquet F. C., Tautz F. S. 2015 Approaching Truly Freestanding Graphene: The Structure of Hydrogen-Intercalated Graphene on 6 H−SiC (0001) *Physical review letters* 114(10):106804

[41] Deretzis I., and La Magna A 2011 Role of covalent and metallic intercalation on the electronic properties of epitaxial graphene on SiC (0001) *Phys Rev B* 84(23): 235426

[42] Forti S., Emtsev K. V., Coletti C., Zakharov A. A., Riedl C., and Starke, U

[58] Tozzini V., and Pellegrini V. 2011 Reversible hydrogen storage by controlled buckling of graphene layers *J Phys. Chem. C* 115(51):25523–25528

[59] Patchkovskii S., John S. T., Yurchenko S. N., Zhechkov L., Heine T., and Seifert G. 2005 Graphene nanostructures as tunable storage media for molecular hydrogen *Proceedings of the National Academy of Sciences of the United States of America* 102(30):10439-10444

[60] Garberoglio G., Pugno N. M., and Taioli S. 2015 Gas adsorption and separation in realistic and idealized frameworks of organic pillared graphene: a comparative study *The Journal of Physical Chemistry C* 119(4):1980-1987

[61] Baburin I. A., Klechikov A., Mercier G., Talyzin A., and Seifert G. 2015 Hydrogen adsorption by perforated graphene *International journal of hydrogen energy* 40(20):6594-6599

[62] Pérez-Jordá J., and Becke A. D. 1995 A density-functional study of van der Waals forces: rare gas diatomics *Chemical physics letters* 233(1):134-137

[63] Kristyán S., and Pulay P. 1994 Can (semi) local density functional theory account for the London dispersion forces? *Chemical physics letters* 229(3):175-180

[64] Miura, Y., Kasai, H., Diño, W., Nakanishi, H., and Sugimoto, T. 2003 First principles studies for the dissociative adsorption of H2 on graphene. *J. Appl. Phys.* 93:3395−3400

[65] Dzhurakhalov, Abdiravuf A., and Francois M. Peeters. "Structure and energetics of hydrogen chemisorbed on a single graphene layer to produce graphane." Carbon 49.10 (2011): 3258-3266.

[66] Wehling T. O., Katsnelson M. I., and Lichtenstein A. I. 2009 Impurities on graphene: Midgap states and migration barriers *Phys Rev B* 80(8):085428

[67] Zhao L., He R., Rim K. T., Schiros T., Kim K. S., Zhou H., Gutiérrez C., Chockalingam S. P., Arguello C. J., Pálová L., Nordlund D., Hybertsen M. S., Reichman D. R., Heinz T. F., Kim P., Pinczuk A., Flynn G. W., and Pasupathy A. N. 2011 Visualizing individual nitrogen dopants in monolayer graphene *Science* 333(6045):999-1003

[68] Kim B. H., Hong S. J., Baek S. J., Jeong Y. H., Park N., Lee M., Lee S. W., Park M., Chu S. W., Shin H. S., Lim J., Lee J. C., Jun Y., and Park Y. W. 2012 N-type graphene induced by dissociative $H_2$ adsorption at room temperature Scientific Reports 2:690

[69] Gallouze M., Kellou A., and Drir M. 2016 Adsorption isotherms of $H_2$ on defected graphene: DFT and Monte Carlo studies *Int. J. Hydrog. Energy* 41:5522–5530

[70] Seydou M., Lassoued K., Tielens F., Maurel F., Raouafi F., and Diawara B. 2015 A DFT-D Study of Hydrogen Adsorption on Functionalized Graphene *RSC Adv*. 5:14400-14406

[71] Ao Z. M., Peeters F. M. 2010 Electric field: A catalyst for hydrogenation of graphene Applied Physics Letters 96(25):253106

[72] Shi S., Hwang J-Y, Li X., Sun X., and Lee B. I. 2010 Enhanced hydrogen sorption on carbonaceous sorbents under electric field *International Journal of Hydrogen Energy* 35(2):629–631

[73] Ao Z. M., Hernández-Nieves A. D., Peeters F. M., and Li S. 2012 The electric field as a novel switch for uptake/release of hydrogen for storage in nitrogen doped graphene *Phys. Chem. Chem. Phys.* 14(4):1463-1467

[74] Ao Z.; and Li S. 2011 Hydrogenation of Graphene and Hydrogen Diffusion Behavior on Graphene/Graphane Interface. *In Graphene Simulation*; Gong, J. R., Ed.; Intech; Chapt. 4:53–74

[75] Ao Z., and Li S. 2014 Electric field manipulated reversible hydrogen storage in graphene studied by DFT calculations *Phys. Status Solidi A* 211(2):351-356

[76] Costanzo, F., Silvestrelli, P. L., and Ancilotto, F. 2012 Physisorption, Diffusion, and Chemisorption Pathways of $H_2$ Molecule on Graphene and on (2,2) Carbon Nanotube by First Principles Calculations. *J. Chem. Theory Comput.* 8, 1288–1294

[77] Gao X., Wang Y., Liu X., Chan T.L., Irle S., Zhao Y., and Zhang S.B. 2011 Regioselectivity control of graphene functionalization by ripples *Phys Chem Chem Phys*. 13(43):19449-53

[78] Boukhvalov D. W., Katsnelson M. I., 2009 Enhancement of Chemical Activity In Corrugated Graphene. *J. Phys. Chem. C* 113:14176− 14178

[79] McKay, H.; Wales, D. J.; Jenkins, S. J.; Verges, J. A.; and de Andres, P. L. 2010 Hydrogen on graphene under stress: Molecular dissociation and gap opening. *Phys. Rev. B* 81:075425

[80] Boukhvalov D. W., Katsnelson M. I., and Lichtenstein A. I. 2008 Hydrogen on graphene: Electronic structure, total energy, structural distortions and magnetism from first-principles calculations. *Physical Review B* 77(3):035427

[81] Leenaerts O., Partoens B., and Peeters F. M. 2008 Adsorption of H 2 O, N H 3, CO, N O 2, and NO on graphene: A first-principles study *Physical Review B* 77(12):125416

[82] Wehling T. O., Katsnelson M. I., and Lichtenstein A. I. 2009 Adsorbates on graphene: Impurity states and electron scattering *Chemical Physics Letters* 476(4):125-134

[83] Chan K. T., Neaton J. B., and Cohen M. L. 2008 First-principles study of metal adatom adsorption on graphene *Physical Review B* 77(23):235430

[84] Takahashi K, S Isobe, K Omori, T Mashoff, D Convertino, V Miseikis, C Coletti, V Tozzini, and S Heun 2016. Revealing the Multibonding State between Hydrogen and Graphene-Supported Ti Clusters *J. Phys. Chem. C*, 120:12974–12979

[85] Parambhath V. B., Nagar R., Sethupathi K., and Ramaprabhu S. 2011 Investigation of Spillover Mechanism in Palladium Decorated Hydrogen Exfoliated Functionalized Graphene *J. Phys. Chem. C* 115(31):15679–15685

[86] Gierz I., Riedl C., Starke U., Ast C. R., and Kern K. 2008 Atomic hole doping of graphene *Nano letters* 8(12): 4603-4607

[87] Dai J., Yuan J., and Giannozzi P. 2009 Gas adsorption on graphene doped with B, N, Al, and S: a theoretical study *Applied Physics Letters* 95(23):232105

[104] Lindsay L., and Broido D. A. 2010 Optimized Tersoff and Brenner empirical potential parameters for lattice dynamics and phonon thermal transport in carbon nanotubes and graphene *Phyl Rev B* 81(20):205441
[105] Tersoff J. 1988 Empirical interatomic potential for carbon, with applications to amorphous carbon *Phys Rev Let* 61(25):2879
[106] Mortazavi B., Rajabpour A., Ahzi S., and Hadizadeh kheirkhah A. 2012 Non-Equilibrium Molecular Dynamics Study on the Thermal and Mechanical Properties of Graphene, Proceedings of the 4th International Conference on Nanostructures (ICNS4) 12-14 March, 2012, Kish Island, I.R. Iran, p. 1255
[107] Magnin Y., Förster G. D., Rabilloud F., Calvo F., Zappelli A., and Bichara C. 2014 Thermal expansion of free-standing graphene: benchmarking semi-empirical potentials *J Phys: Cond Mat*, 26(18):185401
[108] Lu Q., Arroyo M., and Huang R. 2009 Elastic bending modulus of monolayer graphene *Journal of Physics D: Applied Physics* 42(10):102002
[109] Huang Y., Wu J., and Hwang K. C. 2006 Thickness of graphene and single-wall carbon nanotubes *Physical review B* 74(24):245413
[110] Lu Q., and Huang R. 2012 Mechanical Behavior of Monolayer Graphene by Continuum and Atomistic Modeling *in "Simulations in Nanobiotechnology"* Taylor & Francis, p. 485
[111] Lampin E., Priester C., Krzeminski C., and Magaud L. 2010 Graphene buffer layer on Si-terminated SiC studied with an empirical interatomic potential *Journal of Applied Physics* 107(10):103514
[112] Tang C., Meng L., Xiao H., and Zhong J. 2008 Growth of graphene structure on 6H-SiC (0001): Molecular dynamics simulation *Journal of Applied Physics* 103(6):063505